\documentclass[12pt]{article}
\usepackage{epsfig, amssymb}
\usepackage{graphicx,epsfig}
\usepackage{color} 
\begin{document}
\thispagestyle{empty}
\begin{flushright}
\end{flushright}

\bigskip

\begin{center}
\noindent{\Large \textbf
{Electric-magnetic duality as a quantum operator and more symmetries of $U(1)$ gauge theory 
}}\\ 
\vspace{2cm} \noindent{Hobin Lee${}^{a}$\footnote{e-mail:hobin7946@naver.com}, Sanghoon Han${}^{a}$\footnote{e-mail:oksk0729@naver.com}, Hyein Yoon${}^{a}$\footnote{e-mail:lokkun@hanyang.ac.kr}, Junsoo Kim${}^{a}$\footnote{e-mail:rhyeul0103@gmail.com} and 
Jae-Hyuk Oh${}^{a}$\footnote{e-mail:jack.jaehyuk.oh@gmail.com}}

\vspace{1cm}
  {\it
Department of Physics, Hanyang University, Seoul 133-791, Korea${}^{a}$\\
 }
\end{center}

\vspace{0.3cm}
\begin{abstract}
We promote the Noether charge of the electric-magnetic duality symmetry of $U(1)$ gauge theory, ``$G$" to a quantum operator. We construct ladder operators, $D_{(\pm)a}^\dagger(k)$ and $D_{(\pm)a}(k)$ which create and 
annihilate the simultaneous quantum eigen states of the quantum Hamiltonian(or number) and the electric-magnetic duality  operators respectively. Therefore all the quantum states of the $U(1)$ gauge fields can be expressed by a form of $|E,g\rangle$, where $E$ is the energy of the state, the $g$ is the eigen value of the quantum operator $G$, where the $g$ is quantized in the unit of 1. We also show that 10 independent bilinears comprised of the creation and annihilation operators can form $SO(2,3)$ which is as demonstrated in the Dirac's paper published in 1962. The number operator and the electric-magnetic duality operator are the members of the $SO(2,3)$ generators. We note that there are two more generators which commute with the number operator(or Hamiltonian). We prove that these generators are indeed symmetries of the $U(1)$ gauge field theory action. 

\end{abstract}

\section*{Motivation and summary}
Maxwell equations without any electric sources enjoy an interesting symmetry called electric-magnetic duality symmetry\cite{Calkin}. Under a transformation of $(\vec E,\vec B)\rightarrow (\vec B, -\vec E)$, the Maxwell equations are invariant.
In fact, the symmetry can be realized in the form of the (infinitesimal) canonical transformation, which is given by
\begin{equation}
\delta \vec E=\theta \nabla \times \vec A {\rm \ \ and\ \ }\delta \vec A=\theta \nabla^{-2}\nabla \times \vec E,
\end{equation} 
where the $\theta$ is an infinitesimal rotation angle\cite{Deser1}. Such a symmetry appears not even in Maxwell theory but in linearized Einstein gravity,\cite{{Henneaux:2004jw}}, Bosonic and Fermionic gauge field theories\cite{Deser:2004xt}, partially massless (gravity) theories\cite{Deser:2013xb}, and as an approximate symmetries in non-Abelian gauge theories in a few different contexts\cite{Deser:2005sz,Jatkar:2012mm}. In the real world, electric-magnetic duality symmetry is not respected since there are no magnetic monopoles once considering interactions with charged matter. However, the symmetry can be approximately restored by experiment in a certain material system\cite{Fernandez-Corbaton:2013yba}.

This symmetry is also able to be realized as a rotation of the electric and magnetic couplings, which is called ``{\it S-duality}.''\cite{Dijkgraaf:1997ip,Witten:1995gf,Verlinde:1995mz} It is widely known that the symmetry group is SL(2,$\mathcal R$), but if the electric and magnetic charges are introdced and then it is broken down to SL(2,$\mathcal Z$) by Dirac's quantization condition. In this note, however, we discuss the symmetry in the Maxwell theory without charges.

Noether theorem implies that there is a corresponding conserved charge due to the electric-magnetic duality symmetry. It is given by
\begin{equation}
\label{ODG}
G=\frac{1}{2}\int d^3 x[\vec E \cdot \nabla^{-2}\vec \nabla \times \vec E -\vec A \cdot\vec \nabla \times \vec A ].
\end{equation}
Because this is a symmetry generator, $\{H,G\}=0$, where $H$ is the Hamiltonian, given by
\begin{equation}
{ H(\vec E,\vec A)}=\frac{1}{2}\int d^3 x[(\vec E)^2+ (\vec \nabla \times \vec A)^2 ]
\end{equation}
and $\{C,D\}=\frac{\delta C}{\delta A_a}\frac{\delta D}{\delta E_a}-\frac{\delta C}{\delta E_a}\frac{\delta D}{\delta A_a}$ is the Poison bracket. The $\vec E$ represents the Electric field and the $\vec A$ is the vector potential. 
The Maxwell action that we discuss is 
\begin{equation}
S=\int dt\left[ \int d^3x\vec E(t,x)\cdot \dot{\vec{A}}(t,x) - H(\vec E,\vec A) \right],
\end{equation}
where the ``$\cdot$'' between the two fields represents scalar product between two vectors. The ``$\cdot$'' on top of a field is time derivative.

In this note, we rewrite the vector potential and the electric field in terms of certain creation and annihilation operators, which provides a Hilbert space of the simultaneous Eigen states of the quantum operators of $H$ and $G$. Once one assumes that there is no negative norm state, which gives the quantization of the ``$G$''-charge with unit 1. In fact, the charge ``$G$'' is the generator of the rotation of the polarization\cite{Robert,Agullo:2018iya}.



\section*{Quantization of the $U(1)$ gauge fields with the electric-magnetic duality generator}

In the first part of this note, we quantize the $U(1)$ gauge field theory and construct quantum states labeled by their energy eigen values and electric magnetic duality charge. The traditional way to quantize the classical field theory is that one solves the classical field equations, find positive and negative frequency modes and promotes the coefficient of each mode to annihilation and creation operators satisfying a certain commutation relation between them. If one does follow this standard process, then it is found that the creation or annihilation operators are indeed not the ladder operators for the electric-magnetic duality generator, $G$ even though they are the ones for the Hamiltonian.

We start with definition of the creation and annihilation operators, which are given by
\begin{equation}
A_a(k)=\frac{1}{\sqrt{2|k|}}(\mathcal A_a(k)+\mathcal A^\dagger_a(k)), {\ \ \rm and\ \ }E_a(k)=-i\sqrt{\frac{|k|}{2}}(\mathcal A_a(k)-\mathcal A^\dagger_a(k)),
\end{equation}
where $A_a$ and $E_a$ are the gauge fields and the electric fields respectively and $\mathcal A_a$ and $\mathcal A^\dagger_a$ are the creation and the annihilation operators in momentum space. The index $a$ runs over 1 to 3, and so $k_a$ are 3-momentum.

To construct the simultaneous eigen states of the quantum operators of $H$ and $G$, we introduce the other creation and annihilation operators, which are given by  
\begin{figure}[b!]
\centering
\includegraphics[width=120mm]{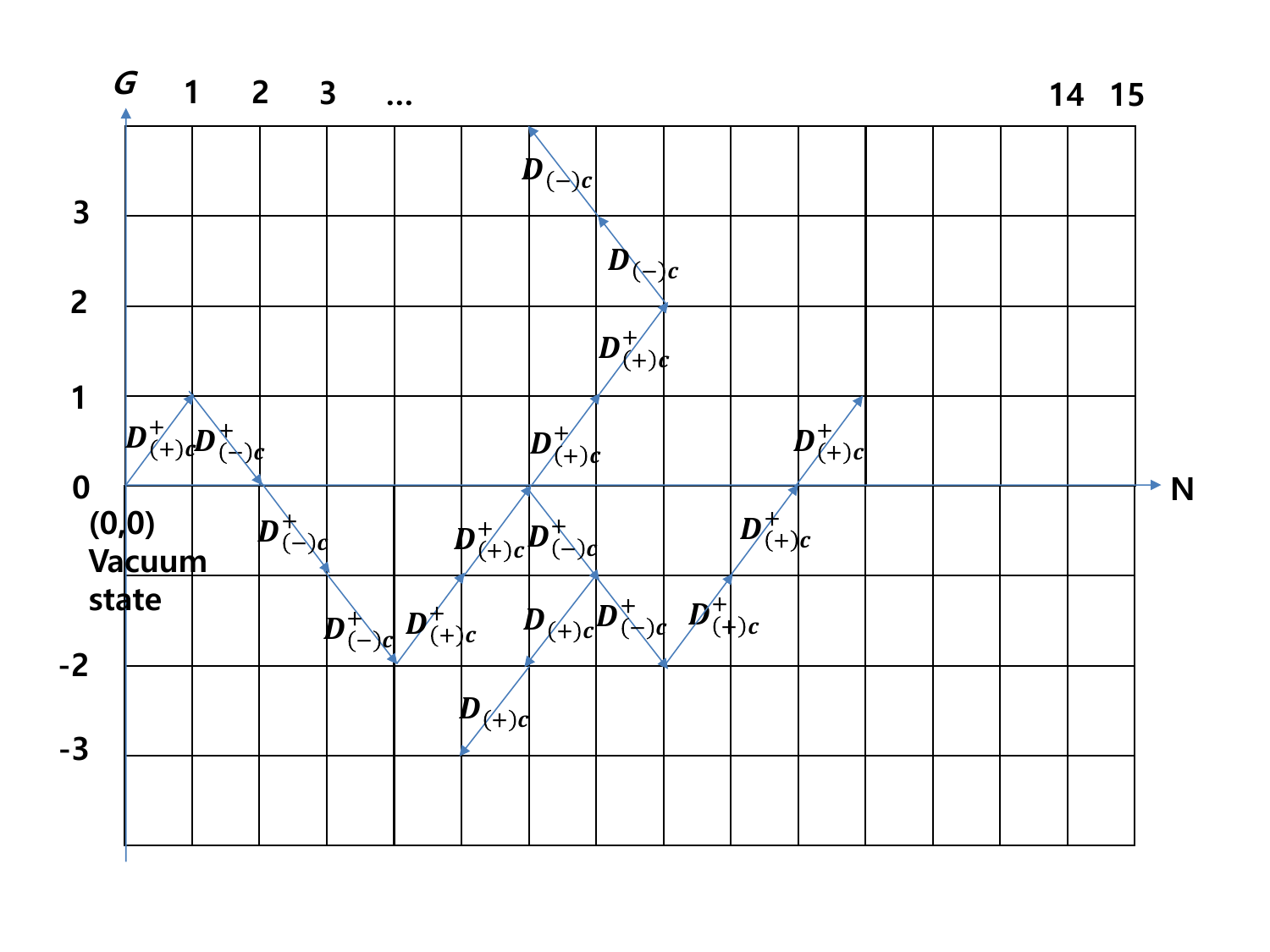}
\caption{The map of the quantum states. The $x$-axis is the eigen-value of the number operator and the $y$-axis is the one of the electric-magnetic duality operator.}
\label{fig1}
\end{figure}
\begin{eqnarray}
D^\dagger_{(\mp)c}(k)\equiv\mathcal A^\dagger_c(k)\pm i\epsilon_{abc}\frac{k_b}{|k|}\mathcal A^\dagger_a(k), \\
D_{(\mp)c}(k)\equiv\mathcal A_c(k)\mp i\epsilon_{abc}\frac{k_b}{|k|}\mathcal A_a(k).
\end{eqnarray}
The new operators have the following properties: $D^\dagger_{(\mp)c}(k)$ are the creation operators in a sense that they increase the energy of the states with amount of $|k|$, but $D^\dagger_{(+)c}(k)$ increase the electric-magnetic duality charge with a unit of 1 whereas $D^\dagger_{(-)c}(k)$ decreases that in the same amount. 
Likewise, $D_{(\mp)c}(k)$ are the annihilation operators for the Hamiltonian but they change the electric-magnetic duality charge with amount of $\pm$1 respectively. 
Therefore, the quantum states are labeled by the two quantum numbers, energy, $E$ and the electric-magnetic duality charge $g$, as $|E,g\rangle$.

Interesting properties of the states are listed in order. Firstly, the vacuum state  is $g=0$ state. Therefore, we express the vacuum as $|0,0\rangle$. Secondly, for the $N$-particle eigen states of $H$ and $G$, $|E,g\rangle$, if $N$ is even, then $g$ is an even number and if $N$ is odd, then $g$ is an odd number.

\paragraph{The emergent $SO(2,3)$ from the bilinears of the $D^\dagger_{(\pm)c}(k)$ or $D_{(\pm)c}(k)$} 
The second issue that we deal with is the bilinear operators and their group structure. We comprise of 10 independent bilinear operators made out of the $D_{(\mp)c}(k)$ and $D^\dagger_{(\mp)c}(k)$ operators. The 4 of them are give by
\begin{eqnarray} 
S_{3}&=&\frac{1}{16} \sum_{\xi=+,-}\int d^{3}k (D^{\dagger}_{(\xi)a}(k)D_{(\xi)a}(k)+D_{(\xi)a}(k)D^{\dagger}_{(\xi)a}(k))=\frac{1}{2}\hat N\\ 
\nonumber
&=&\frac{1}{4} \int d^{3}k (\mathcal A^{\dagger}_{a}(k)\mathcal A_{a}(k)+\mathcal A_{a}(k)\mathcal A^{\dagger}_{a}(k)) \\
L_{2}&=&-\frac{1}{8} \int d^{3}k (D^{\dagger}_{(+)a}(k)D_{(+)a}(k)-D^{\dagger}_{(-)a}(k)D_{(-)a}(k)) 
=-\frac{1}{2} \hat{G} \\ \nonumber
&=&-\frac{1}{2i}\int d^{3}k \epsilon_{efa}\frac{k_f}{|k|}\mathcal A^{\dagger}_{e}(k) \mathcal A_{a}(k) \\
K_{2}&=&\frac{i}{8} \int d^{3}k (D^{\dagger}_{(+)a}(k)D^{\dagger}_{(-)a}(k)-D_{(+)a}(k)D_{(-)a}(k)) \\ \nonumber
&=&\frac{i}{4} \int d^{3}k (\mathcal A^{\dagger}_{a}(k)\mathcal A^{\dagger}_{a}(k)-\mathcal A_{a}(k)\mathcal A_{a}(k)) \\
Q_{2}&=&-\frac{1}{8} \int d^{3}k (D^{\dagger}_{(+)a}(k)D^{\dagger}_{(-)a}(k)+D_{(+)a}(k)D_{(-)a}(k)) \\ \nonumber
&=&-\frac{1}{4}\int d^{3}k (\mathcal A^{\dagger}_{a}(k)\mathcal A^{\dagger}_{a}(k)+\mathcal A_{a}(k)\mathcal A_{a}(k)) 
\end{eqnarray}
These operators are commute with the electric-magnetic duality generator, $G$. The electric-magnetic duality generator in terms of the primitive annihilation and creation operators
\footnote{It is manifest that the expression of $G$ in terms of the annihilation and creation operators (\ref{GITOACOP}) does not divergent when $|k|\rightarrow 0$, where the $\frac{k^a}{|k|}$ is unit vector in the momentum, $k^a$ direction, which is finite in any case. In the original definition of $G$ given in (\ref{ODG}), (\ref{ODGKK}) in position and momentum spaces respectively, they seems to be divergent in that limit. In fact, it is not. In Lagrangian formulation, the electric field $\vec E^T=\omega \vec A^T$ and for a real photon, $\omega=|k|$. The $\vec E^T$, $\vec A^T$ are the transverse parts of the elecctric and gauge fields respectively, which are the genuine degrees of freedom. Thus, $\frac{E^T}{|k|}$ is finite as $|k|\rightarrow 0$, where $\omega$ is energy of photon(See also \cite{IWO}). One can interpret the operator $G$ is helicity operator and quantization of its Eigen values with unit $1$ is reasonable in that sense. The original definition of helicity is given by $\vec L \cdot \frac{\vec k}{|k|}$, where the $\vec L$ is angular momentum of photons. It is a projection of the photon angular momentum along the direction of the propagation, which is finite by definition\cite{Calkin}.}
is given by
\begin{equation}
\label{GITOACOP}
G\equiv {-}{i}\int d^3k \frac{\epsilon_{abc}k_b }{|k|} \mathcal A^\dagger_a(k) \mathcal A_c(k).
\end{equation}
By employing commutation relations between $G$ and $\mathcal A^\dagger_a(k)$, $ \mathcal A_c(k)$, one can realize that $G$ is a $SO(2)$ rotation generator and  $\vec\mathcal A=$($\mathcal A_1(k),\mathcal A_2(k)$) and $\vec \mathcal A^\dagger=$($\mathcal A^\dagger_1(k),\mathcal A^\dagger_2(k)$) are vectors under such a transform, where we set the direction of momentum of the gauge fields to be $\vec k=(0,0,k)$. In fact, they transform as
\begin{equation}
\left(\begin{array}{cc}
   \mathcal A_1(k) \\
   \mathcal A_2(k) \\
  \end{array}\right)^\prime
  =\left(\begin{array}{cc}
   \cos \theta & \sin \theta \\
   -\sin \theta & \cos \theta \\
  \end{array}\right)
  \left(\begin{array}{cc}
   \mathcal A_1(k) \\
   \mathcal A_2(k) \\
  \end{array}\right),
\end{equation}
and the creation operators do in the same way.
It is manifest that the above 4 bilinear operators are invariant under the $SO(2)$ rotation since they are either inner products or fully anti symmetric combination of the vector components of $\vec \mathcal A$ or $\vec \mathcal A^\dagger$. Therefore, they are commute one another.

The symmetric tensor part of the bilinear combination of the vectors $\vec \mathcal A$ or $\vec \mathcal A^\dagger$ are given in the below.
\begin{eqnarray}\nonumber
L_{1}&=&\frac{1}{8}\int d^{3}k [D^{\dagger}_{+(1}(k)D_{2)-}(k)+D^{\dagger}_{-(1}(k)D_{2)+}(k)]\\ \nonumber
L_{3}&=&-\frac{i}{8}\int d^{3}k [D^{\dagger}_{+(1}(k)D_{2)-}(k)-D^{\dagger}_{-(1}(k)D_{2)+}(k)]\\ \nonumber
K_{1}&=&\frac{i}{16}\int d^{3}k [D^{\dagger}_{+(1}(k)D^{\dagger}_{2)+}(k)-D^{\dagger}_{-(1}(k)D^{\dagger}_{2)-}(k)+D_{-(1}(k)D_{2)-}(k)-D_{+(1}(k)D_{2)+}(k)]\\ \nonumber
K_{3}&=&\frac{1}{8}\sum_{\xi=+,-}\int d^{3}k [D^{\dagger}_{\xi(1}(k)D^{\dagger}_{2)\xi}(k)+D_{\xi(1}(k)D_{2)\xi}(k)]\\ \nonumber
Q_{1}&=&-\frac{1}{16}\int d^{3}k [D^{\dagger}_{+(1}(k)D^{\dagger}_{2)+}(k)-D^{\dagger}_{-(1}(k)D^{\dagger}_{2)-}(k)-D_{-(1}(k)D_{2)-}(k)+D_{+(1}(k)D_{2)+}(k)]\\ \nonumber
Q_{3}&=&\frac{i}{8}\sum_{\xi=+,-}\int d^{3}k [D^{\dagger}_{\xi(1}(k)D^{\dagger}_{2)\xi}(k)-D_{\xi(1}(k)D_{2)\xi}(k)]\\ \nonumber
\end{eqnarray}
These bilinear operators are the symmetric traceless components of $SO(2)$ tensors as $\mathcal A_a \mathcal A^\dagger_b$, $\mathcal A^\dagger_a \mathcal A^\dagger_b$ or $\mathcal A_a \mathcal A_b$. Absolutely these will change under the $SO(2)$ rotation and consequently they do not commute with $G$ i.e. $L_2$. The commutators between the bilinear operators are given in the table below.

\begin{table}[hbp]
\centering
\begin{tabular}{|c|c|c|c|c|c|c|c|c|c|c|}
\hline
 &$L_1$&$L_2$&$L_3$&$S_3$&$K_1$&$K_2$&$K_3$&$Q_1$&$Q_2$&$Q_3$ \\ 
\hline
$L_1$&0&$i L_3$ &$-i L_2$ &0 &0 &$i K_3$ &$-i K_2$ &0 &$i Q_3$ &$-i Q_2$ \\
\hline
$L_2$&$-i L_3$ &0 &$i L_1$ &0 &$-i K_3$ &0 &$i K_1$ &$-i Q_3$ &0 &$i Q_1$ \\
\hline
$L_3$&$i L_2$ &$-i L_1$ & 0 & 0 &$ i K_2$&$-i K_1$ & 0 &$ i Q_2$ &$-i Q_1$ & 0 \\
\hline
$S_3$&0&0& 0 &0 &$ -i Q_1$ &$ -i Q_2$ &$-i Q_3$&$i K_1$&$ i K_2$ &$ i K_3$ \\
\hline
$K_1$&0&$i K_3$&$-i K_2$ & $ i Q_1$&0 &$-i L_3$&$i L_2$&$i S_3$&0&0 \\
\hline
$K_2$&$-i K_3$ &0 & $ i K_1$&$ i Q_2$ &$i L_3$ &0 &$-{\color{red}i}L_1$ &0 &$i S_3$ &0 \\
\hline
$K_3$&$i K_2$ &$-i K_1$ & 0 & $ i Q_3$&$-i L_2$ &${\color{red}i}L_1$ &0 &0 &0 &$i S_3$ \\
\hline
$Q_1$&0 &$i Q_3$ &$-i Q_2$ &$-i K_1$ &$-i S_3$ &0 &0 &0 &$-i L_3$ &$ i L_2$ \\
\hline
$Q_2$&$-i Q_3$ &0 &$ i Q_1$ &$ -i K_2$ &0 &$-i S_3$ &0 &$ i L_3$ &0 &$-i L_1$ \\
\hline
$Q_3$&$i Q_2$ &$-i Q_1$ & 0 &$ -i K_3$ &0 &0 &$-i S_3$ &$-i L_2$ & $i L_1$&0 \\
\hline 
\end{tabular}
\caption{Values of commutator , [column , row ]}
\end{table}
In fact, the tensor components transform under the $SO(2)$ rotation as
\begin{eqnarray} 
L'_1&=&L_{1}\cos({2\theta})+L_{3} \sin({2\theta}) \\ 
L'_3&=&-L_{1} \sin({2\theta})+L_{3}\cos({2\theta}) \\ 
K'_1&=&K_{1}\cos({2\theta})+K_{3} \sin({2\theta})   \\ 
K'_3&=&-K_{1} \sin({2\theta}) +K_{3}\cos({2\theta}) \\ 
Q'_1&=&Q_{1}\cos({2\theta})+Q_{3}  \sin({2\theta}) \\ 
Q'_3&=&-Q_{1} \sin({2\theta})+Q_{3}\cos({2\theta})
\end{eqnarray}
This means that the combinations of the tensor components as $L^2_{1}+L^2_{2}$, $K^2_{1}+K^2_{2}$ and $Q^2_{1}+Q^2_{2}$ are invariant under electric-magnetic duality rotation.


\section*{More electric-magnetic duality like symmetries} 
Once one observes the $SO(2,3)$ bilinear generators, one may realize that the operators $L_1$ and $L_3$  commute with the number operator(also with the Hamiltonian) as well as the $L_2$, which is the electric-magnetic duality generator. This means that $L_1$ and $L_3$ are candidates of the symmetry generator of the $U(1)$ gauge theory. The Hamiltonian is obtained by Legendre transformation from the Lagrangian. Then once a quantity, $\int d^4x \vec E(x,t) \cdot \vec A(x,t)$ is invariant under $L_1$ or $L_3$, the Lagrangian will be so too. In the following discussion, we formulate the symmetry transformation in the language of $SO(2)$ rotation. This is due to that the operators, $L_1$ and $L_2$ are the tensor components of the $SO(2)$ rotation along the direction of the momentum, $\vec k$, which is generated by the $G$. we set $\vec k=|k| \hat x_3$, where $\hat x_3$ is the third directional unit vector in the 3-dimensional flat space.

\paragraph{The symmetry generator $L_1$ and its transformation}
To check if they are indeed a symmetry of the $U(1)$ gauge field theory action, let us examine the $L_1$ operator first. We start with the relation between the creation and annihilation operators and the electric fields and the gauge fields, which are given by
\begin{equation}
A_a(k)=\frac{1}{\sqrt{2|k|}}(\mathcal A_a(k)+\mathcal A^\dagger_a(k)), {\ \ \rm and\ \ }E_a(k)=-i\sqrt{\frac{|k|}{2}}(\mathcal A_a(k)-\mathcal A^\dagger_a(k)).
\end{equation}
By using the above expression, we get the transformation of the fields $A_a(k)$ and $E_a(k)$ under $L_1$, which are given by

\begin{eqnarray}
\delta A_a&=&\epsilon[L_1,A_a(k)]=-i\frac{\epsilon}{2|k|}\epsilon^+_{3cd}\Pi_{ca}(k)E_{d}(k), \\
\delta E_a&=&\epsilon[L_1,E_a(k)]=i\frac{\epsilon}{2}|k|\epsilon^+_{3cd}\Pi_{ca}(k)A_d(k),
\end{eqnarray}
where $\epsilon^+_{abc}$ are symmetric and off-diagonal symbol defined as
\begin{eqnarray}
\epsilon^+_{abc}&=&0 {\ \ \rm if\ any\ of\ the\ indices\ are\ the\ same\ with\ an(the)\ other(s),} \\
\epsilon^+_{abc}&=&1 {\ \ \rm if\ all\ the\ indices\ are\ different\ one\ another.}
\end{eqnarray}
For instance, $\epsilon^+_{112}=\epsilon^+_{333}=0$, $\epsilon^+_{123}=\epsilon^+_{132}=1$.
It is obvious that the Hamiltonian is invariant under the transformation. Since the action is obtained by Legendre transformation from the Hamiltonian as
\begin{equation}
S=\int d^3kdt E_a(k,t) \dot A_a(k,t) - \int H(E_a,A_a) dt,
\end{equation}
and we understand that the Hamiltonian is invariant under the above transformation, to prove that the action, $S$ is invariant, we need to show that $\int d^3kdt{\ } E_i(k,t) \dot A_i(k,t)$ does not change upto total derivative under it. The change of the term is given by
\begin{eqnarray}
\nonumber
&\delta&\left(\int d^3kdt{\ }E_i(k,t) \dot A_i(k,t)\right)=\int d^3kdt{\ } \delta E_i(k,t) \dot A_i(k,t) + \int d^3kdt{\ } E_i(k,t) \delta\dot A_i(k,t)
\\ \nonumber
&=&\int d^3kdt \left(i\frac{\epsilon}{4}|k|\epsilon^+_{3cd}\Pi_{ca}(k)\frac{\partial (A_d(k) A_a(-k))}{\partial t} -i\frac{\epsilon}{4}\frac{1}{|k|}\epsilon^+_{3cd}\Pi_{ca}(k)
\frac{\partial (E_d(k) E_a(-k))}{\partial t}\right).
\end{eqnarray}
Therefore, it it manifest that the $L_1$ is a symmetry generator.

One probably asks if this can get into the finite version of the transformation by exponentiate the infinitesimal one. 
One realize that the creation and annihilation operators transform as
\begin{equation}
\left( {\begin{array}{cc}
 \mathcal  A_1 \\
 \mathcal A_2 \\
  \end{array} } \right)^\prime
  =
  \left( 
\begin{array}{cc}
   \cosh\frac{\theta}{2} & -\sinh\frac{\theta}{2}  \\
 -\sinh\frac{\theta}{2} &  \cosh\frac{\theta}{2}  \\
 \end{array} 
\right)
\left( \begin{array}{cc}
 \mathcal  A_1 \\
 \mathcal A_2 \\ 
\end{array}
\right),
\end{equation} 
and
\begin{equation}
\left( {\begin{array}{cc}
 \mathcal  A^\dagger_1 \\
 \mathcal A^\dagger_2 \\
  \end{array} } \right)^\prime
  =
  \left( 
\begin{array}{cc}
   \cosh\frac{\theta}{2} & \sinh\frac{\theta}{2}  \\
 \sinh\frac{\theta}{2} &  \cosh\frac{\theta}{2}  \\
 \end{array} 
\right)
\left( \begin{array}{cc}
 \mathcal  A^\dagger_1 \\
 \mathcal A^\dagger_2 \\ 
\end{array}
\right).
\end{equation} 
It is a pseudo rotation between them. The annihilation operators transform with its rotation angle $\theta$ whereas the creation operators transformation is with $-\theta$. The gauge and electric fields change under the symmetry generator operation as
\begin{equation}
\left( {\begin{array}{cccc}
   A_1 \\
  A_2\\
E_1  \\
E_2\\
  \end{array} } \right)^\prime
  =
  \left( {\begin{array}{cccc}
   \cosh\frac{\theta}{2} & 0 & 0& -\frac{i}{|k|}\sinh\frac{\theta}{2} \\
  0&   \cosh\frac{\theta}{2} & -\frac{i}{|k|}\sinh\frac{\theta}{2}& 0 \\
0 & i|k|\sinh\frac{\theta}{2}&\cosh\frac{\theta}{2} &  0 \\
i|k|\sinh\frac{\theta}{2}& 0& 0 & \cosh\frac{\theta}{2}   \\
  \end{array} } \right)
\left( {\begin{array}{cccc}
   A_1 \\
  A_2\\
E_1  \\
E_2\\
  \end{array} } \right).
\end{equation}

Under such a transformation, the Hamiltonian(the number operator) is invariant manifestly. What matter is the transform of the term $\int \vec E(k)\cdot \dot A(k)$ in the action. This changes as
\begin{eqnarray}
\int \vec E^\prime(k)\cdot \dot {\vec A^\prime}(k) d^3kdt \rightarrow \int \vec E(k)\cdot \dot {\vec A}(k) d^3kdt 
\\ \nonumber
+\int d^3k dt \frac{\partial}{\partial t}\left \{ \sinh^2\frac{\theta}{2} (A_1 E_1+A_2 E_2) + \sinh\frac{\theta}{2}\cosh\frac{\theta}{2}\left(i|k| A_1A_2 - \frac{i}{|k|}E_1E_2 \right)\right \}
\end{eqnarray}.


\paragraph{The symmetry generator $L_3$ and its transformation}
The 2nd and last operator that we examine is the  $L_3$ operator. The transformation of the fields $A_a(k)$ and $E_a(k)$ when we act this operator on them is given by
\begin{eqnarray}
\delta A_a&=&\epsilon[L_3,A_a(k)]=-i\frac{\epsilon}{2}\frac{\Pi^-_{ab}(k)}{|k|}E_{b}(k), \\
\delta E_a&=&\epsilon[L_3,E_a(k)]=i\frac{\epsilon}{2}|k|\Pi^-_{ab}(k)A_b(k),
\end{eqnarray}
where $\Pi^-_{ab}(k)$ is given by
\begin{equation}
\Pi^-_{ab}(k)=-\epsilon_{3ca}\epsilon^+_{3cb}
\end{equation}
The Hamiltonian is invariant under the transformation. Again, we show that $\int d^3kdt{\ } E_i(k,t) \dot A_i(k,t)$ does not change upto total derivative under this transform as
\begin{eqnarray}
\nonumber
&\delta&\left(\int d^3kdt{\ }E_i(k,t) \dot A_i(k,t)\right)=\int d^3kdt{\ } \delta E_i(k,t) \dot A_i(k,t) + \int d^3kdt{\ } E_i(k,t) \delta\dot A_i(k,t)
\\ \nonumber
&=&\int d^3kdt \left(i\frac{\epsilon}{4}|k|\Pi^-_{ab}(k)\frac{\partial (A_a(k) A_b(-k))}{\partial t} -i\frac{\epsilon}{4}\frac{\Pi^-_{ab}(k)}{|k|}
\frac{\partial (E_a(k) E_b(-k))}{\partial t}\right)
\end{eqnarray}


One probably asks if this can get into the finite version of the transformation by exponentiate the infinitesimal one. 
One realize that the creation and annihilation operators transform as
\begin{equation}
\left( {\begin{array}{cc}
 \mathcal  A_1 \\
 \mathcal A_2 \\
  \end{array} } \right)^\prime
  =
  \left( 
\begin{array}{cc}
   e^{-\frac{\theta}{2}} & 0 \\
0 &  e^{\frac{\theta}{2}} \\
 \end{array} 
\right)
\left( \begin{array}{cc}
 \mathcal  A_1 \\
 \mathcal A_2 \\ 
\end{array}
\right),
\end{equation} 
and
\begin{equation}
\left( {\begin{array}{cc}
 \mathcal  A^\dagger_1 \\
 \mathcal A^\dagger_2 \\
  \end{array} } \right)^\prime
  =
  \left( 
\begin{array}{cc}
   e^{\frac{\theta}{2}} & 0 \\
0 &  e^{-\frac{\theta}{2}} \\
 \end{array} 
\right)
\left( \begin{array}{cc}
 \mathcal  A^\dagger_1 \\
 \mathcal A^\dagger_2 \\ 
\end{array}
\right).
\end{equation} 
It is a kind of chiral rotation between them. 
The creation operators rotate with the oppositie angle againest the annihilation operators.
The gauge and electric fields change under the symmetry generator operation as
\begin{equation}
\left( {\begin{array}{cccc}
   A_1 \\
  A_2\\
E_1  \\
E_2\\
  \end{array} } \right)^\prime
  =
  \left( {\begin{array}{cccc}
   \cosh\frac{\theta}{2} & 0 &  -\frac{i}{|k|}\sinh\frac{\theta}{2}& 0 \\
  0&   \cosh\frac{\theta}{2} & 0 & \frac{i}{|k|}\sinh\frac{\theta}{2}  \\
i|k|\sinh\frac{\theta}{2} & 0 &\cosh\frac{\theta}{2} &  0 \\
0 & -i|k|\sinh\frac{\theta}{2} & 0 & \cosh\frac{\theta}{2}   \\
  \end{array} } \right)
\left( {\begin{array}{cccc}
   A_1 \\
  A_2\\
E_1  \\
E_2\\
  \end{array} } \right).
\end{equation}

Under such a transformation, the Hamiltonian(the number operator) is invariant manifestly. What matter is the transform of the term $\int \vec E(k)\cdot \dot A(k)$ in the action. This changes as
\begin{eqnarray}
\int \vec E^\prime(k)\cdot \dot {\vec A^\prime}(k) d^3kdt \rightarrow \int \vec E(k)\cdot \dot {\vec A}(k) d^3kdt 
\\ \nonumber
+\int d^3k dt \frac{\partial}{\partial t}\left \{ \sinh^2\frac{\theta}{2} (A_1 E_1+A_2 E_2) + \sinh\frac{\theta}{2}\cosh\frac{\theta}{2}\left(i|k| \frac{A^2_1-A^2_2}{2} - \frac{i}{|k|}\frac{E^2_1-E^2_2}{2} \right)\right \}
\end{eqnarray}.

\section*{Discussion}
In this note, we discuss electric-magnetic duality symmetry. Its generator is a member of a group $SO(2,3)$. The group, $SO(2,3)$ is not a symmetry group of Maxwell theory defined in $\mathbb R^{1,3}$ . Under the transformations generated by $L_1$, $L_3$, $K_1$, $K_3$, $Q_1$, and  $Q_3$, the Maxwell action changes. The set of $\{S_3, L_i\}$, where $i=1,2,3$ forms a Cartan subgroup of $SO(2,3)$, which is indeed the symmetry group of the Maxwell theory. This is $SL(2,R)$.

Free Maxwell theory enjoys novel spacetime symmetry, which is conformal symmetry group of $SO(2,4)$ which is ensured by a fact that stress-energy tensor of Maxwell theory is traceless. The Poincare and Lorentz groups, $SO(1,4)\supset SO(1,3)$ are the subgroups of the conformal symmetry group. The Poincare group is isomorphic to the group of $SO(2,3)$ that we obtained in this note taking account of Wick rotation of one of the non-compact direction to compact one. The structure is similar but their physical origins are different.

We also see this as follows. We note that the number operator and electric-magnetic duality operator have definite physical meanings. 
Especially, the electric-magnetic duality operator is helicity operator of photon states. Since Maxwell theory is translational invariant, one may consider the corresponding Noether charges as symmetry generators. There are 4 generatrors, which are the Hamiltonian density and momentum density operators are $\mathcal H=|k|\mathcal A^\dagger_a(k) \mathcal A_a(k)$ and $\vec \mathcal P=\vec k\mathcal A^\dagger_a(k) \mathcal A_a(k)$. However, in the $SO(2,3)$, the symmetric combinations of $\mathcal A^\dagger_a(k)$ and $\mathcal A_a(k)$ with their index summation i.e. $\sum_{a}\mathcal A^\dagger_a(k) \mathcal A_a(k)$ is the number operator only. Therefore, there is no one to one correspondance between spacetime symmetry generators and the $SO(2,3)$ generators.





\section*{Appendices}

\subsection*{A. Quantization with the electric-magnetic duality generator and the quantum operators and states}
The electric-magnetic duality generator in momentum space is given by
\begin{equation}
\label{ODGKK}
G=\frac{i}{2}\int d^3 k \epsilon_{abc}\left[ E_a(k) \frac{k_b}{k^2}  E_c(-k) + A_a(k) k_b A_c(-k) \right],
\end{equation}
where $E_a(k)$ are the electric fields and $A_a(k)$ are the gauge fields. They are the canonical pair and so satisfy the following Poisson bracket relation:
\begin{equation}
\{ A_b(x),E_a(y)\}=\delta_{ab}\delta^{(3)}(x-y)
\end{equation}
The two fields satisfy the Gauss constraint, $\partial_a E^a=0$ and this ensures that they are transverse fields. Therefore the Poisson bracket relation is modified as
\begin{equation}
\{ A^{T}_b(x),E^{T}_a(y)\}=\Pi_{ab}(x)\delta^{(3)}(x-y),
\end{equation}
where the $\Pi_{ab}(x)=\left(\delta_{ab}-\frac{\partial_a \partial_b}{\nabla^2}\right)$ is the project operator.

\paragraph{Quantization}
Maxwell theory is mathematically a collection of two independent harmonic oscillators.

\begin{eqnarray}
A_a(x)=\frac{1}{(2\pi)^{\frac{3}{2}}}\int d^3 k \exp(ik_b x_b)\sum_{A=1,2}q_A(k)e^A_a(k), \\ \nonumber
E_a(x)=\frac{1}{(2\pi)^{\frac{3}{2}}}\int d^3 k \exp(ik_b x_b)\sum_{A=1,2}p^A(k)e^a_A(k),
\end{eqnarray}
where the indices $a,b...$ are 3 dimensional spatial indices and $A,B...$ are the polarization indices.
The $e^a_A(k)$ is the polarization vector,  $e^a_A(k)k_a=0$ to make sure that the fields $A_a(x)$ and $E_a(x)$ are transverse and $e^a_A(k)e^B_a(-k)=\delta^B_A$. $p^A(k)=p^A(-k)^\star$, $q^A(k)=q^A(-k)^\star$ and $e^A_a(k)=e^A_a(-k)^\star$ because the fields are real. $e^A_b(k_1)e^a_A(-k_1)\equiv\Pi^a_b(k_1)=\delta^a_b-\frac{k_ak_b}{k^2}$.

Fourier transform,  
\begin{equation}
\phi(x)=\frac{1}{(2\pi)^{\frac{3}{2}}}\int d^3 k \exp(ik_b x_b)\phi(k),
\end{equation}
defines the fields in the momentum space as
\begin{equation}
A_a(k)=q_A(k)e^A_a(k){\rm \ \ and\ \ }E^a(k)=p^A(k)e_A^a(k).
\end{equation}
Their Poisson brackets are given by
\begin{eqnarray}
\{ A_b(k_1),E^a(k_2)  \}=\Pi_a^b(k_1)\delta^{(3)}(k_1+k_2), \\ \nonumber
\{ q_A(k_1),p^B(k_2)  \}=\delta_A^B\delta^{(3)}(k_1+k_2).
\end{eqnarray}
We define creation and annihilation operators as
\begin{eqnarray}
a_A(k)=\frac{1}{\sqrt{2}}  \left( \sqrt{|k|}q_A(k)+\frac{i}{\sqrt{|k|}}p_A(k) \right), \\ \nonumber
a^\dagger_A(k)=\frac{1}{\sqrt{2}}  \left( \sqrt{|k|}q_A(-k)-\frac{i}{\sqrt{|k|}}p_A(-k) \right),
\end{eqnarray}
and the inverse relations are given by
\begin{eqnarray}
q_A(k)=\frac{1}{\sqrt{2|k|}}(a_A(k)+a^\dagger_A(-k)), \\ \nonumber
p_A(k)=-i\sqrt{\frac{|k|}{2}}(a_A(k)-a^\dagger_A(-k)).
\end{eqnarray}
The final form of the Poisson bracket is given by 
\begin{eqnarray}
\{ \mathcal A^\dagger_d(k),G  \}&=&\epsilon_{abc}\frac{k_b}{|k|}\Pi_{dc}(k)\mathcal A_a^\dagger(k), \\ \nonumber
\{ \mathcal A_d(k),G  \}&=&\epsilon_{abc}\frac{k_b}{|k|}\Pi_{dc}(k)\mathcal A_a(k), \\ \nonumber
\{ \mathcal A_a(k),\mathcal A^\dagger_b(k^\prime)  \}&=&-i\Pi_{ab}(k) \delta^{(3)}(k-k^\prime), \\ \nonumber
\{ H,\mathcal A_a(k)  \}&=& i|k|\mathcal A_a(k), \\ \nonumber
\{ H,\mathcal A^\dagger_a(k)  \}&=&-i|k|\mathcal A^\dagger_a(k),
\end{eqnarray}
where $\mathcal A_a(k)=e^A_a(-k) a_A(k)$ and $\mathcal A^\dagger_a(k)=e^A_a(k) a^\dagger_A(k)$. The quantization of the fields is performed by switching the Poisson brackets to the commutators as $\{{\ \ }\} \rightarrow -i[{\ \ }]$ and promoting all the fields to the quantum operators.
The forms of the Hamiltonian and electric-magnetic duality operators are given by
\begin{equation}
H=\int d^3k |k|a^\dagger_A(k)a^A(k)\equiv\int d^3k |k| \mathcal A^\dagger_a(k)\mathcal A^a(k),
\end{equation}
and
\begin{equation}
G={\color{red}-}{i}\int d^3k \frac{\epsilon_{abc}k_b e^A_a(k)e^B_c(-k)}{|k|} a^\dagger_A(k) a_B(k)\equiv {\color{red}-}{i}\int d^3k \frac{\epsilon_{abc}k_b }{|k|} \mathcal A^\dagger_a(k) \mathcal A_c(k).
\end{equation}
The annihilation and creation operators are ladders for the Hamiltonian but those do not increase or decrease the eigen values of the operator, $G$. To find simultaneous ladders for the $H$ and $G$, we define
\begin{eqnarray}
D^\dagger_{(\mp)c}(k)=\mathcal A^\dagger_c(k)\pm i\epsilon_{abc}\frac{k_b}{|k|}\mathcal A^\dagger_a(k), \\
D_{(\mp)c}(k)=\mathcal A_c(k)\mp i\epsilon_{abc}\frac{k_b}{|k|}\mathcal A_a(k),
\end{eqnarray}
and then it turns out that the commutation relations are modified to
\begin{eqnarray}
[D^\dagger_{(\mp)c}(k),G]=\pm D^\dagger_{(\mp)c}(k), {\ \ \ }[D_{(\mp)c}(k),G]={\color{red}\mp} D_{(\mp)c}(k), \\ \nonumber
[H,D^\dagger_{(\mp)c}(k)]=|k|D^\dagger_{(\mp)c}(k), {\ \ \ }[H,D_{(\mp)c}(k)]=-|k|D_{(\mp)c}(k), \\ \nonumber
[D_{(\mp)c}(k),D^\dagger_{(\mp)d}(k^\prime)]=2\delta^{(3)}(k-k^\prime)\left( \Pi_{cd}(k)\pm i \epsilon_{ced}\frac{k_e}{|k|}\right).
\end{eqnarray}

\section*{B. Construction of $[SO(2,3)]$ group from bilinear operators from $D_{(\pm)a}(k)$ and $D^\dagger_{(\pm)a}(k)$}

We start with a new definition of $D_{(\pm)a}(k)$ and $D^\dagger_{(\pm)a}(k)$ for further conveience, which is given by
\begin{eqnarray}
D^\dagger_{(\xi)c}(k)=\mathcal A^\dagger_c(k)- i\xi\epsilon_{abc}\frac{k_b}{|k|}\mathcal A^\dagger_a(k), \\
D_{(\xi^\prime)c}(k)=\mathcal A_c(k)+ i\xi^\prime\epsilon_{abc}\frac{k_b}{|k|}\mathcal A_a(k), \\ \nonumber
\end{eqnarray}
where for $\xi=\pm 1$, $D^\dagger_{(\xi)c}(k)$ represents $D^\dagger_{(\pm)c}(k)$ and for $\xi^\prime=\pm 1$, $D_{(\xi^\prime)c}(k)$ represents $D_{(\pm)c}(k)$ respectively. The bilinear operators constructed out of the operators, $D_{(\xi)c}(k)$ and $D^\dagger_{(\xi)c}(k)$ have the following forms:
\begin{eqnarray}
 \int D^\dagger_{(\xi)a}(k)D^\dagger_{(\xi')b}(k) d^3 k, \\
 \int D_{(\xi)a}(k)D_{(\xi')b}(k) d^3 k, \\ 
 \int D_{(\xi)a}(k)D^\dagger_{(\xi')b}(k)d^3k, \\
 \int D^\dagger_{(\xi)a}(k)D_{(\xi)b}(k)d^3 k,
\end{eqnarray}
where the third and the fourth operators are the same upto a constant(a c-number). Since we are interested in their commutation relations, we regard the third and fourth as the same. One can also classify the operators into their trace, anti-symmetric and traceless-symmetric parts.

First of all, we discuss their trace parts, which are listed as below:
\begin{eqnarray}
\label{traceunun}
\int d^3 k{\ } D^\dagger_{(\xi)a}(k)D^\dagger_{(\xi')a}(k) =(1-\xi\xi^\prime)\int d^3 k{\ } \mathcal A^\dagger_{a}(k)\mathcal A^\dagger_{a}(k) , \\
\label{tracedd}
\int d^3 k{\ }D_{(\xi)a}(k)D_{(\xi')a}(k) =(1-\xi\xi^\prime)\int d^3 k{\ }\mathcal A_{a}(k)\mathcal A_{a}(k) ,
\end{eqnarray}
and
\begin{equation}
\label{tracedun}
\int d^3 k{\ } D^\dagger_{(\xi)a}(k)D_{(\xi')a}(k)=(1+\xi\xi')\int d^3 k{\ } \mathcal A^\dagger_{a}(k)\mathcal A_{a}(k)-i(\xi+\xi')\int d^3 k{\ } \epsilon_{efa}\frac{k_f}{|k|}\mathcal A^\dagger_{e}(k)\mathcal A_{a}(k).\\ \nonumber
\end{equation}
When $\xi$ and $\xi^\prime$ take the same sign, the bilinear operators (\ref{traceunun}) and (\ref{tracedd}) become null identically. If they take the different sign as $(\xi,\xi^\prime)=(+1,-1)$ or $(\xi,\xi^\prime)=(-1,+1)$, then (\ref{traceunun}) and (\ref{tracedd})are proportional to the trace of the primitive creation and annihilation operators respectively. Since the operators are not Hermitian, we constitute their appropriate linear combinations to be Hermitian operators. They are nothing but $K_2$ and $Q_2$ operators listed below.

\begin{eqnarray}  
K_{2}&=&\frac{i}{8} \int d^{3}k (D^{\dagger}_{(+)a}(k)D^{\dagger}_{(-)a}(k)-D_{(+)a}(k)D_{(-)a}(k)) \\ \nonumber
&=&\frac{i}{4} \int d^{3}k (\mathcal A^{\dagger}_{a}(k)\mathcal A^{\dagger}_{a}(k)-\mathcal A_{a}(k)\mathcal A_{a}(k)) \\ 
Q_{2}&=&-\frac{1}{8} \int d^{3}k (D^{\dagger}_{(+)a}(k)D^{\dagger}_{(-)a}(k)+D_{(+)a}(k)D_{(-)a}(k)) \\ \nonumber
&=&-\frac{1}{4}\int d^{3}k (\mathcal A^{\dagger}_{a}(k)\mathcal A^{\dagger}_{a}(k)+\mathcal A_{a}(k)\mathcal A_{a}(k)) \\ \nonumber
\end{eqnarray}

Only when $\xi$ and $\xi^\prime$ take the same sign, (\ref{tracedun}) becomes non-trivial, which is a linear combination of the number operator and electric-magnetic duality operator. Together with this, we examine the anti-symmetric parts of the bilinear operators, which are given by
\begin{eqnarray} 
\label{antidd}
\int d^{3}k{\ }D^\dagger_{(\xi)[a}D^\dagger_{b](\xi')}=i(-\xi+\xi')\int d^{3}k{\ }\frac{k_f}{|k|}(-\epsilon_{efb}\mathcal A^\dagger_{a}\mathcal A^\dagger_{e}+\epsilon_{efa}\mathcal A^\dagger_{b}\mathcal A^\dagger_{e})\\ 
\label{antiunun}
\int d^{3}k{\ }D_{(\xi)[a}D_{b](\xi')}=i(-\xi+\xi')\int d^{3}k{\ }\frac{k_f}{|k|}(\epsilon_{efb}\mathcal A_{a}\mathcal A_{e}-\epsilon_{efa}\mathcal A_{b}\mathcal A_{e})\\ 
\label{antidun}
\int d^{3}k{\ }D^\dagger_{(\xi)[a}D_{b](\xi')}=\int d^{3}k{\ }\left[\mathcal A^\dagger_{a}\mathcal A_{b}-\mathcal A^\dagger_{b}\mathcal A_{a}+\xi\xi'\frac{k_dk_f}{|k|}\mathcal A^\dagger_{c}\mathcal A_{e}(\epsilon_{cda}\epsilon_{efb}-\epsilon_{cdb}\epsilon_{efa})\right.\\ \nonumber
+\left.\epsilon_{efb}\frac{k_f}{|k|}(i\xi'\mathcal A^\dagger_{a}\mathcal A_{e}+i\xi\mathcal A^\dagger_{e}\mathcal A_{a})-\epsilon_{efa}\frac{k_f}{|k|}(i\xi'\mathcal A^\dagger_{b}\mathcal A_{e}+i\xi\mathcal A^\dagger_{e}\mathcal A_{b})\right]
\end{eqnarray}
It turns out that (\ref{antidd}) and (\ref{antiunun}) are not independent operators. Once we contract the operators with a tensor $\frac{k_c}{|k|}\epsilon_{abc}$, they become proportional to (\ref{traceunun}) and (\ref{tracedd}) respectively. The same contractions acting on (\ref{antidun}) leads another linear combination of the number and electric-magnetic duality operators. Therefore, appropriate combinations of the (\ref{antidun}) and (\ref{tracedun}) provide the operators below.

\begin{eqnarray}
\nonumber
S_{3}&=&\frac{1}{16} \int d^{3}k (D^{\dagger}_{(+)a}(k)D_{(+)a}(k)+D^{\dagger}_{(-)a}(k)D_{(-)a}(k)+D_{(-)a}(k)D^{\dagger}_{(-)a}(k)+D_{(+)a}(k)D^{\dagger}_{(+)a}(k))\\ \nonumber
&=&\frac{1}{4} \int d^{3}k (\mathcal A^{\dagger}_{a}(k)\mathcal A_{a}(k)+\mathcal A_{a}(k)\mathcal A^{\dagger}_{a}(k))=\frac{1}{2}\hat N \\ \nonumber
L_{2}&=&-\frac{1}{8} \int d^{3}k (D^{\dagger}_{(+)a}(k)D_{(+)a}(k)-D^{\dagger}_{(-)a}(k)D_{(-)a}(k)) \\ \nonumber
&=&-\frac{1}{2i}\int d^{3}k \epsilon_{efa}\frac{k_f}{|k|}\mathcal A^{\dagger}_{e}(k) \mathcal A_{a}(k)=-\frac{1}{2} \hat{G} 
\end{eqnarray}

To construct the symmetric parts of the bilinear operators, we utilize the following identities: 
\begin{eqnarray}\nonumber
D^\dagger_{(\xi)(a}D^\dagger_{b)(\xi')}=2\mathcal A^\dagger_{a}\mathcal A^\dagger_{b}-2\xi \xi' \epsilon_{cdb}\epsilon_{efa}\frac{k_dk_f}{|k|^2}\mathcal A^\dagger_{c}\mathcal A^\dagger_{e}-i(\xi+\xi')\frac{k_f}{|k|}(\epsilon_{efb}\mathcal A^\dagger_{a}\mathcal A^\dagger_{e}+\epsilon_{efa}\mathcal A^\dagger_{b}\mathcal A^\dagger_{e})\\ \nonumber
D_{(\xi)(a}D_{b)(\xi')}=2\mathcal A_{a}\mathcal A_{b}-2\xi \xi' \epsilon_{cdb}\epsilon_{efa}\frac{k_dk_f}{|k|^2}\mathcal A_{c}\mathcal A_{e}+i(\xi+\xi')\frac{k_f}{|k|}(\epsilon_{efb}\mathcal A_{a}\mathcal A_{e}+\epsilon_{efa}\mathcal A_{b}\mathcal A_{e})\\ \nonumber
D^\dagger_{(\xi)(a}D_{b)(\xi')}=\mathcal A^\dagger_{a}\mathcal A_{b}+\mathcal A^\dagger_{b}\mathcal A_{a}+\xi\xi'\frac{k_dk_f}{|k|}\mathcal A^\dagger_{c}\mathcal A_{e}(\epsilon_{cda}\epsilon_{efb}+\epsilon_{cdb}\epsilon_{efa})\\ \nonumber+\epsilon_{efb}\frac{k_f}{|k|}(i\xi'\mathcal A^\dagger_{a}\mathcal A_{e}-i\xi\mathcal A^\dagger_{e}\mathcal A_{a})+\epsilon_{efa}\frac{k_f}{|k|}(i\xi'\mathcal A^\dagger_{b}\mathcal A_{e}-i\xi\mathcal A^\dagger_{e}\mathcal A_{b})
\end{eqnarray}
Since they are tensor components, they change under $SO(3)$ spatial rotation. Therefore, we choose our frame as $\vec k=(0,0,k)$ and then the spatial index $a$ in $D_{a(\xi)}(k)$ and $D^\dagger_{a(\xi)}(k)$ can take $1$ or $2$. Every possible (linearly independent one another) choices are listed below.
\begin{eqnarray}\nonumber
L_{1}&=&\frac{1}{8}\int d^{3}k [D^{\dagger}_{+(1}(k)D_{2)-}(k)+D^{\dagger}_{-(1}(k)D_{2)+}(k)]\\ \nonumber
L_{3}&=&-\frac{i}{8}\int d^{3}k [D^{\dagger}_{+(1}(k)D_{2)-}(k)-D^{\dagger}_{-(1}(k)D_{2)+}(k)]\\ \nonumber
K_{1}&=&\frac{i}{16}\int d^{3}k [D^{\dagger}_{+(1}(k)D^{\dagger}_{2)+}(k)-D^{\dagger}_{-(1}(k)D^{\dagger}_{2)-}(k)+D_{-(1}(k)D_{2)-}(k)-D_{+(1}(k)D_{2)+}(k)]\\ \nonumber
K_{3}&=&\frac{1}{8}\int d^{3}k [D^{\dagger}_{+(1}(k)D^{\dagger}_{2)+}(k)+D^{\dagger}_{-(1}(k)D^{\dagger}_{2)-}(k)+D_{-(1}(k)D_{2)-}(k)+D_{+(1}(k)D_{2)+}(k)]\\ \nonumber
Q_{1}&=&-\frac{1}{16}\int d^{3}k [D^{\dagger}_{+(1}(k)D^{\dagger}_{2)+}(k)-D^{\dagger}_{-(1}(k)D^{\dagger}_{2)-}(k)-D_{-(1}(k)D_{2)-}(k)+D_{+(1}(k)D_{2)+}(k)]\\ \nonumber
Q_{3}&=&\frac{i}{8}\int d^{3}k [D^{\dagger}_{+(1}(k)D^{\dagger}_{2)+}(k)+D^{\dagger}_{-(1}(k)D^{\dagger}_{2)-}(k)-D_{-(1}(k)D_{2)-}(k)-D_{+(1}(k)D_{2)+}(k)]\\ \nonumber
\end{eqnarray}

\section*{Acknowledgement}
J.H.O thanks his $\mathcal {W.J.}$ and $\mathcal {D.D}$. He also thank Hyun Seok Yang for useful discussion. This work was supported by the National Research 
Foundation of Korea(NRF) grant funded by the Korea government(MSIP) 
(No.2016R1C1B1010107) and Research Institute for Natural Sciences, Hanyang University.

\end{document}